\documentclass{elsart}
\usepackage{graphicx,amssymb}
\journal{Physics Letters B}
\begin{document}
\begin{frontmatter}


\title{The flavor problem and discrete symmetries}
\author{A. C. B. Machado\thanksref{ana}},
\ead{ana@ift.unesp.br}
\author{V. Pleitez\thanksref{vp}}
\address{Instituto de F\'\i sica Te\'orica,
Universidade Estadual Paulista\\ Rua Pamplona 145,
01405-900 - S\~ao Paulo, SP, Brazil
}
\ead{vicente@ift.unesp.br}
\thanks[ana]{Supported by CNPq}
\thanks[vp]{Partially supported by CNPq under the process
300613/2005-9}



\begin{abstract}
In this Letter we propose a multi-Higgs extension of the standard model with Abelian
and non-Abelian discrete symmetries in which the mass matrices of the charged fermions
obtained from renormalizable interactions are diagonal. However, non-diagonal contributions, 
that are important for obtaining the CKM matrix in the quark sector, 
arise from non-renormalizable dimension five interactions. Active neutrinos acquire mass only from 
non-renormalizable interactions, the non-diagonal entries arising through dimension five operators, 
while the diagonal entries comes from dimension six operators. Realistic mixing matrices in the 
neutrino and the quarks sectors are obtained.
\end{abstract}
\end{frontmatter}


The flavor problem can be divided in several subproblems~\cite{volkas}: Why do weak
isospin partners have different masses? Why are quark and lepton masses split? Why is
there a mass hierarchy between generations? Presently we know that neutrino oscillation
data are well described by massive neutrinos~\cite{concha} and the flavor problem becomes
more interesting:
why is there a mixing angle hierarchy in the quark sector but not in the lepton sector?
Usually mass matrices have the form $M_{\alpha\beta}=\sum_i(\Gamma_i)_{\alpha\beta} \langle
\Phi^0_i\rangle$, where the $\Gamma_i$'s are, for Dirac fermions, arbitrary complex
dimensionless $3\times 3$ matrices, and $\langle\Phi^0_i\rangle$ denotes one or a set of vacuum
expectation values (VEVs) of the neutral scalar fields in the model. For Majorana fermions,
the $\Gamma_i$'s are complex symmetric matrices. The mixing matrix and the mass pattern of each
charged sector depend on the structure of the respective $\Gamma_i$'s. It is well known that
explicit and predictive forms of these matrices can be obtained by imposing flavor
symmetries. For instance, it has been considered global unitary (horizontal) symmetries like
$U(1)$'s, $SU(2)_H$~\cite{zeewilczek} or $SU(3)_H$~\cite{koide}. More recently symmetries
like $S_n$, $Z_n$~\cite{zhang}; $A_4$~\cite{a4}, $^{(d)}T$~\cite{phf1,tprime}, the double
covering of $A_4$; $Q_6$, the double dihedral group~\cite{babu}, and other discrete
symmetries~\cite{phf2} have been also considered.

Here we will turn the problem upside down. Mass matrices will be of the form $M_{\alpha\beta}=
f(\overline{\langle\Phi^0\rangle})_{\alpha\beta}$ in which $f$ denotes a few dimensionless
($\mathcal{O}(1)$) parameters (one at leading order), and $\overline{\langle\Phi^0\rangle}$ is
a matrix built with the VEVs of several scalar fields. At first sight, there is no gain in
predictive power, we are just changing a dimensionless general matrix $\Gamma_{\alpha\beta}$ by
another one with mass dimension $(\overline{\langle\Phi^0\rangle})_{\alpha\beta}$. However, it
seems easier, at least in principle, to explain patterns of dynamical variables like VEVs, than
dimensionless numbers. The value of the former can be explained by the dynamics (for instance by
studying the scalar potential) and extra flavor symmetries that we can impose to the model.

In this Letter we will consider an extension  of the standard model in which the charged
fermion masses, that arise from renormalizable interactions, are diagonal. However,
dimension 5 operators~\cite{dim5} give important non-diagonal contributions to quarks but not
to the charged leptons. On the other hand, the neutrinos mass matrix is generated mainly from  
dimension 5 and 6 operators, with the former having zero trace as in the Zee's mode ~\cite{zee1}.
The dimension 6 operators induce non-negligible diagonal terms and a general $3 \times
3$ symmetric mass matrix for those neutrinos is obtained. The non-diagonal elements are 
important in the quark sector because they involve also the VEV of the singlet which it is not 
constrained by the electroweak scale, and also in the neutrino sector because theses entries 
involve ratios of doublets' VEVs that produce an enhancement. This enhancement does not 
occurs in the charged lepton sector because in this case the non-diagonal entries depend on the
same doublet's VEVs that appear in the diagonal ones but are suppressed by $\Lambda^{-1}$. 
No right-handed neutrinos are introduced.
In summary, we show that all the questions mentioned above have a natural answer: fermion masses
and mixing arise from different scalar sectors and from different effective interactions. The
predictive power is a consequence of the discrete symmetries imposed to the model. 

Simple non-Abelian discrete groups have been considered for the first time as candidates for
flavor symmetries by Frampton and Kephart~\cite{phf1}. Later, $A_4$ symmetry has been used to
obtain realistic mixing matrices in the lepton sector~\cite{ma1}, and then also in the quark
sectors~\cite{ma2}. More recently~\cite{a4tribi}, by using these non-Abelian discrete symmetries
it was possible to obtain a tribimaximal neutrino mixing matrix~\cite{tribi}.  The basic idea is
to consider $SU(2)_{L}$ fermion doublets and singlets in the triplet representation of the $A_4$
symmetry and Higgs multiplets transforming as triplet or singlet of $A_4$.

Let us briefly review some properties of the $A_4$ representations.
Denoting the vector basis ($\textbf{3}$) as $(a_1,a_2,a_3)$ and $(b_1,b_2,b_3)$ we have
$\textbf{3}\otimes \textbf{3}=\textbf{3}_A\oplus \textbf{3}_B\oplus \textbf{1}
\oplus \textbf{1}^\prime\oplus \textbf{1}^{\prime\prime}$. Explicitly~\cite{he},
\begin{eqnarray}
&&[\textbf{3}\otimes \textbf{3}]_A\equiv\textbf{3}_A =(a_2b_3, a_3b_1,a_1b_2),\quad
[\textbf{3}\otimes \textbf{3}]_B\equiv \textbf{3}_B =(a_3b_2, a_1b_3,a_2b_1), \nonumber \\
&&[\textbf{3}\otimes \textbf{3}]_1=a_1b_1+a_2b_2+a_3b_3,\quad\quad\;\;
[\textbf{3}\otimes \textbf{3}]_{1^\prime}=a_1b_1+\omega a_2b_2+\omega^2a_3b_3,\nonumber \\ &&
[\textbf{3}\otimes \textbf{3}]_{1^{\prime\prime}}=a_1b_1+\omega^2a_2b_2+\omega a_3b_3,
\label{produtos}
\end{eqnarray}
and $\textbf{1}\otimes \textbf{1}=\textbf{1}$, $\textbf{1}^\prime\otimes
\textbf{1}^{\prime\prime}=\textbf{1}$, $\textbf{1}^\prime\otimes
\textbf{1}^\prime=\textbf{1}^{\prime\prime}$, $\textbf{1}^{\prime\prime}\otimes
\textbf{1}^{\prime\prime}=\textbf{1}^\prime$.

We consider a model with $G_{SM}\otimes A_4 \otimes Z_3 \otimes Z^{\prime}_3 \otimes
Z^{\prime\prime}_3$ symmetry, where $G_{SM}$ is the gauge symmetry $SU(3)_C\otimes 
SU(2)_L\otimes U(1)_Y$. The matter fields transform as usual under $G_{SM}$, left-handed lepton 
and quark doublets $L_a,\,a=e, \mu,\tau$; $Q_i,\,i=1,2,3$, and their respective right-handed 
singlets $l_{aR}$, $u_{iR}$ and $d_{iR}$. In the scalar sector, we introduce scalar doublets with 
$Y=+1$:  $H_i,\hat{H}_i,H^{\prime}_{i}, H^{\prime\prime}_i$, $\Phi_{i}$, $\Phi^{\prime}_{i}$, 
$\Phi^{\prime\prime}_i$ and $\chi$; three scalar triplets with $Y=+2$, denoted by 
$\mathcal{T}_i$~\cite{chengli}, and finally one complex (with $Y=0$) scalar singlet $\zeta$. 
Fermion weak eigenstates fields transform under $(A_4,Z_3,Z^{\prime}_3,Z^{\prime\prime}_3)$ as 
follows:
\begin{eqnarray}
&&L\equiv (L_e,L_\mu,L_\tau) \sim (\textbf{3}, \omega, \omega^2, 1);\quad
\;\;l_R\equiv(\mu_R,\tau_R,e_R) \sim (\textbf{3},1, 1, \omega),\nonumber \\ &&
Q_{L} \equiv (Q_{1L},Q_{2L},Q_{3L}) \sim (\textbf{3}, \omega^2, 1, 1);\;\;\quad
 u_R\equiv (c_R,t_R,u_R) \sim (\textbf{3}, 1, 1, \omega); \nonumber \\ &&
d_R\equiv (s_R,b_R,d_R) \sim (\textbf{3},1, \omega, \omega),
\label{fermions}
\end{eqnarray}
in which $\omega=e^{2\pi i/3}$.  Notice that all fermion fields are still symmetry eigenstates 
but we do not use an special notation. On the other hand, scalar fields transform under
$(A_4,Z_3,Z^\prime_3,Z^{\prime\prime}_3)$ as:
\begin{eqnarray}
&&H\equiv (H_1,H_2,H_3)\sim (\textbf{3},\omega, \omega^2, 1),\;\; \hat{H}\equiv
(\hat{H}_1,\hat{H}_2,\hat{H}_3)\sim (\textbf{3},\omega, \omega^2,  \omega^2),
\nonumber \\ &&
H^\prime\equiv (H^\prime_1,H^\prime_2,H^\prime_3)\sim(\textbf{3}, \omega, 1,  \omega), \;\;
H^{\prime\prime}\equiv (H^{\prime\prime}_1,H^{\prime\prime}_2,
H^{\prime\prime}_3)\sim(\textbf{3}, \omega^{2}, \omega^{2}, \omega^2),\;
\nonumber \\ &&
\Phi\equiv (\Phi_1,\Phi_2,\Phi_3) \sim (\textbf{3}, 1, 1, 1),\;
\Phi^{\prime} \equiv (\Phi^\prime_1,\Phi^\prime_2,\Phi^\prime_3)\sim (\textbf{3},\omega, \omega,
\omega), \nonumber \\ && \;
\Phi^{\prime \prime}\equiv (\Phi^{\prime\prime}_1,\Phi^{\prime\prime}_2,
\Phi^{\prime\prime}_3) \sim (\textbf{3}, \omega^{2}, 1,  \omega^2),\;
 \chi\sim (\textbf{1}, \omega^{2}, \omega, 1),\nonumber \\ &&
 \mathcal{T}\equiv (\mathcal{T}_1,\mathcal{T}_2,\mathcal{T}_3)\sim (\textbf{3},
 \omega^{2}, \omega, 1), \;
  \zeta \sim (\textbf{1}, 1, \omega,1).
 \label{scalars}
\end{eqnarray}

With the fields in (\ref{fermions}) and (\ref{scalars}) we have the leading contributions to the 
lepton and quark Yukawa interactions:
\begin{eqnarray}
\mathcal{L}& = &\frac{1}{\Lambda}\left( f_\nu\,[(\overline{L^c}\epsilon H)]_A
[( L\epsilon \Phi)]_B
+ \frac{f^\prime_\nu}{\Lambda}
[\overline{L^c} \varepsilon\vec{\sigma}
\cdot \vec{\mathcal{T}}]_B[L\Phi^\dagger ]_B\chi+\dots \right)+ H.c.
\nonumber \\ &+&
\left( g_l [\overline{L} \hat{H}]_A + \frac{g^\prime_l}{\Lambda^2} [\overline{L} \hat{H}]_A
\vert\zeta\vert^2 + \cdots \right)
l_R+H.c.
\nonumber \\  &+&
\left( h_u [\overline{Q_L} \widetilde{H}^\prime]_A  + \frac{h^\prime_u}{\Lambda}
[\overline{Q_L}
\widetilde{\Phi}^{\prime} ]_B \zeta +\cdots\right)u_R +H.c.
\nonumber \\  &+& \left(h_d [\overline{Q_L} H^{\prime\prime}]_A    +
\frac{h^\prime_d}{\Lambda}
 [\overline{Q_L} \Phi^{\prime\prime}]_B \zeta^* +\cdots\right)d_R+H.c.,
\label{lagrangiana}
\end{eqnarray}
where $\Lambda$ is an energy scale characterizing an unknown physics; and
$\varepsilon=i\sigma_2$ ($\sigma_2$ being the Pauli matrix), $[~]_{A,B}$ means the appropriate 
product defined in Eq.~(\ref{produtos}) and $\cdots$ denotes higher-dimensional operators, 
we assume that $\vert V\vert\ll \Lambda$, where $V$ denotes the VEV of  any of the scalar of 
the model. 

The contributions to the quark mass matrices obtained from renormalizable interactions 
are diagonal, but the non-renormalizable dimension 5 interactions  
contribute with important and not negligible non-diagonal elements, on  
the other hand, for the charged leptons this corrections come from  
dimension 6 operators are negligible. 
We obtain diagonal entries to the mass matrices as a consequence of the definition of the basis 
in which the product $[[~]_A\cdot]_1$ gives the diagonal contribution while  the product 
$[[~]_B\cdot]_1$ gives the non-diagonal contributions to the mass matrix. For instance, for the 
$u$-quarks, the leading diagonal mass matrix arises through the interaction $[\bar{Q}H^\prime]_A
=(\bar{Q}_2 H^\prime_3,\bar{Q}_3 H^\prime_1,\bar{Q}_1H^\prime_2)$ and then $[[\bar{Q}H^\prime]_A
(c_R,t_R,u_R)]_1=\bar{Q}_2 H^\prime_3c_R+\bar{Q}_3 H^\prime_1t_R+\bar{Q}_1H^\prime_2u_R$. 
Similarly, using the $[~]_B$ product, we obtain that $[[\bar{Q}\Phi^\prime]_B(c_R,t_R,u_R)]_1=
\bar{Q}_3 \Phi^\prime_2c_R+\bar{Q}_1\Phi^\prime_3t_R+\bar{Q}_2 \Phi^\prime_1 u_R$, which gives 
the non-diagonal contributions to the charged lepton mass matrix.

Hence, from the Yukawa interactions in Eq.~(\ref{lagrangiana}), the mass matrices obtained are
\begin{equation}
M^l \approx g_l\left(
\begin{array}{ccc}
\hat{v}_2 & 0 & 0\\
0&\hat{v}_3 & 0 \\
0& 0& \hat{v}_1\end{array}
\right) + H.c.,
\label{leptons}
\end{equation}
for the charged leptons. We have denoted $\langle\hat{H}_i\rangle=\hat{v}_i$. Notice that the 
renormalizable interactions are the dominant one, i.e., $g_l\hat{v_1}\simeq m_\tau$, 
$g_l\hat{v}_3~\simeq~m_\mu$ and $g_l\hat{v}_2\simeq~m_e$. In the quark sector we have
\begin{eqnarray}
M^u \approx h_u\,\left(
\begin{array}{ccc}
v^\prime_2 & 0& a_{u} v^{\prime}_{\phi_3} \\
a_{u} v^{\prime}_{\phi_1} & v^\prime_3 & 0\\
0& a_{u} v^{\prime}_{\phi_2}& v^\prime_1\end{array}\right)
+H.c.,
\label{us}
\end{eqnarray}
and
\begin{equation} 
M^d \approx h_d\,\left(\begin{array}{ccc} v^{\prime\prime}_2 &
0& a_{d} v^{\prime \prime}_{\phi_3} \\ a_{d} v^{\prime \prime}_{\phi_1} &
v^{\prime\prime}_3 & 0 \\ 0& a_{d} v^{\prime \prime}_{\phi_2}&
v^{\prime\prime}_1\end{array}\right)
+ H.c.,
\label{ds}
\end{equation}
in which $a_{u} = \frac{h^\prime_u}{h_{u}} \frac{v_{\zeta}}{\Lambda}$ and $a_{d} =
\frac{h^\prime_d}{h_{d}} \frac{v_{\zeta}}{\Lambda}$,  for the $2/3$ and $-1/3$ charged
quarks, respectively. We have denoted $\langle H^0_i\rangle=v_i$, $\langle H^{\prime0}_i
\rangle=v^\prime_i$, $\langle H^{\prime\prime0}_i\rangle=v^{\prime\prime}_i$,
$\langle\Phi_{i}\rangle=v_{\phi_{i}}$, $\langle\Phi^\prime_i\rangle=v^\prime_{\phi_{i}}$,
$\langle\Phi^{\prime\prime}_{i}\rangle=v^{\prime\prime}_{\phi_{i}}$,
and $\langle\zeta \rangle=v_\zeta$. Notice, as well, that each charged sector has its
private VEVs, thus all these matrices are independent from each other. In the case of
neutrinos we have the mass matrix
\begin{equation}
M^\nu \approx \left[\left(
            \begin{array}{ccc}
             0&
             \frac{v_2}{v_1} & \frac{v_{\phi_3} }{ v_{\phi_1} } \\
              \frac{v_2}{v_1}  &   0 &
              \frac{v_3}{v_1}
              \frac{v_{\phi_2} }{ v_{\phi_1} }  \\
               \frac{v_{\phi_3} }{ v_{\phi_1} } & \frac{v_3}{v_1}
               \frac{v_{\phi_2} }{ v_{\phi_1} }
               & 0 \\
            \end{array}
          \right)+\left(\begin{array}{ccc}
          \delta_1 &
             0 & 0 \\
             0  &   \delta_2 & 0  \\
             0 & 0 & \delta_3 \\
          \end{array}\right)\right]\frac{f_\nu v_{\phi_1} }{ \Lambda } \,v_1,
          \label{neutrinos}
\end{equation}
where the $\delta_i$ are given by
$\delta_1=\frac{f^\prime_\nu}{f_\nu}\,(v_{_{T3}}v_{\phi_3}v_\chi/\Lambda v_1v_{\phi_1})$,
$\delta_2=\frac{f^\prime_\nu}{f_\nu}\,(v_{_{T1}} v_\chi/ \Lambda v_1)$, and $\delta_3=
\frac{f^\prime_\nu}{f_\nu}\,(v_{_{T2}} v_{\phi_2} v_\chi/ \Lambda v_1v_{\phi_1})$,
where $\langle \chi\rangle=v_\chi$, and $\langle\Delta^0_i\rangle=v_{_{Ti}}$.

It is interesting that the discrete symmetries of the model implies that the
charged lepton mass matrix, Eq.~(\ref{leptons}), is practically diagonal, and that the
quark mass matrices have three zeros, see Eqs.~(\ref{us}) and (\ref{ds}).
To the best of our knowledge, this mass matrices have not been considered in
literature~\cite{xing}. Notice that in the case of the neutrino we have already taken
into account the hermitian conjugate.

Since the only constraint on the VEVs of the doublets is that $\sum v^2<
(\textrm{174}\textrm{GeV})^2$ (there are triplets as well, see below), 
a possibility is that only $v^{\prime\,2}_1\lesssim (174\, \textrm{GeV})^2$, 
while all the other VEVs are $\ll 174$ GeV. In this case $v^\prime_1$ give the main 
contribution to the mass of the $t$ quark.  On the other hand, the sum of the VEVs of the triplets has 
to have a upper limit of the order of a few GeV to not spoil the observed value of the $\rho$ 
parameter~\cite{triplet}. Hence, $\vert V\vert/\Lambda$ are well-defined
expansion parameters. The present model is natural in the Yukawa interactions i.e.,
unlike the standard model, there is no a hierarchy among the Yukawa couplings. In practice, in the
analysis below we will consider that all Yukawa coupling are in the interval $0.1-3$. 
The VEV of the singlet $v_\zeta$ is only constrained by the condition 
$v_\zeta/\Lambda< 1$. Below we will use always, just as an illustration: $\Lambda=1$ TeV and 
$v_\zeta=140$ GeV. However, recall that the singlet $\zeta$ may be related to a new energy scale different 
from the electroweak scale.

As a first illustration, let us consider the matrix in Eq.~(\ref{neutrinos}).
We will assume, without losing generality, that this matrix
is a traceless matrix i.e., that the condition $\sum_i\delta_i=0$  is valid~\cite{zee1}.
This is obtained by imposing (using $\tilde{v}_x~=~v_xe^{i\theta_x}$, where $v_x$
and $\theta_x$ are real):
\begin{eqnarray}
&&\theta_{_{T3}}+\theta_{\phi_3}-\theta_1-\theta_{\phi_1}=\pi,\;\; \theta_{_{T1}}-
\theta_1=0;
\;\; \theta_{_{T2}}+\theta_{\phi_1}
-\theta_1-\theta_{\phi_1}=0,\nonumber \\ &&
(1/2)\left\vert\frac{f^\prime_\nu}{f_\nu}\,\frac{ v_{_{T_3}}
v_\chi v_{\phi_3} } {\Lambda v_1 v_{\phi_1}}\right\vert=
\left\vert\frac{f^\prime_\nu}{f_\nu}\,\frac{v_{_{T_1}} v_\chi }
{\Lambda v_1}\right\vert =
\left\vert\frac{f^\prime_\nu}{f_\nu}\,\frac{v_{_{T_2}} v_\chi  v_{\phi_2} }
{\Lambda v_1 v_{\phi_1} }\right\vert \equiv \delta.
\label{ufa}
\end{eqnarray}
Hence, the neutrino mass matrix is of the form~\cite{brama1,brama2}
\begin{equation}
M^\nu\approx m_0\left( \begin{array}{ccc}
-2\delta & \sin\theta & \cos\theta \\
\sin\theta & \delta & \epsilon \\
\cos\theta & \epsilon & \delta
\end{array}\right),
\label{bra1}
\end{equation}
where $m_0=(f_\nu/\Lambda)\vert v_{\phi_1} v_1\vert,\, \vert v_2/v_1 \vert=
\sin\theta, \, \vert v_{\phi_3}/v_{\phi_1} \vert =\cos\theta, \, \epsilon= \vert
(v_3v_{\phi_2}/v_1v_{\phi_1}) \vert$.  From Table~I of Ref.~\cite{brama2} we can see that
neutrinos masses and mixing angles are in agreement with all experimental data if, for
instance, in  Eq.~(\ref{bra1}), $m^2_0=\Delta m^2_{13}\equiv \Delta m^2_{atm}=(0.039\,
\textrm{eV})^2$, $\epsilon=0.21$, $\delta=0.236$, and $\theta=0.84$. In our model, these 
values for the above parameters are obtained by assuming: $\Lambda=1$ TeV, $v_1\sim1$ MeV, 
$v_{\phi_1}=v_{\phi_2}$, $2v_{T_1}=2v_{T_2}=v_{T_3}$. with $v_{T_1}=1.25$ GeV; then 
we obtain: $f_\nu v_{\phi_1}\sim40$ keV, $v_{\phi_3}=40/f_\nu$ keV, $v_2=0.1$ keV, $v_3=0.2$ 
MeV, $(f_\nu/f^\prime_\nu)v_\chi=0.2$ GeV. We also obtain with these VEVs, $\Delta 
m^2_{12}=9\times10^{-5}\,\textrm{eV}^2$,
$\sin^22\theta_\odot=0.82$, $\sin^2 2\theta_{atm}=0.99$, and $\sin\theta_{13}=0.019$ a
value that is in agreement with the CHOOZ limit $V_{e3}=\sin\theta_{13}<0.07$~\cite{chooz}. 
We see that a realistic leptonic mixing matrix arises since the mass matrix of the charged leptons 
is almost diagonal. If $f_\nu, f^\prime_\nu\sim\mathcal{O}(1)$, it implies that some VEVs are of 
the order of MeV or even less depending on the values of $v_\zeta$ and $\Lambda$. 
This does not necessarily implies the existence of light scalars since 
for having them in a model, there must be some symmetries that allow them to appear in the mass 
spectra: The symmetries in the present model do not play that role and $\mu^2_{ij}$ terms
survive in the scalar mass matrices (see Ref.~\cite{ma3}).

On the other hand, with $M^{u,d}$ given by Eq.~(\ref{us}) and (\ref{ds}) we see that $M^{u(d)}M^{u(d)\dagger}
\not=M^{u(d)\dagger}M^{u(d)}$. Hence we need four matrices, $V^{U,D}_{L,R}$, for diagonalizing 
these mass matrices by biunitary transformations and the constraint 
$V_{CKM}=V^U_LV^{D\dagger}_L$. Besides, the matrices $V^{U,D}_{L,R}$ must satisfy the constraints
$V^D_LM^dM^{d\dagger}V^{D\dagger}_L=diag(m^2_d,m^2_s,m^2_b)=(\hat{M}^d)^2$ and 
$V^D_RM^{d\dagger}M^{d}V^{D\dagger}_R=(\hat{M}^d)^2$; similarly for the $V^U_{L,R}$ matrices.
The quark masses and the $V_{CKM}$ matrix arise, for instance, using for the $d$-type quarks 
$a_d=0.14$, $h_d= h^\prime_d=0.1$, and (all VEVs and masses below are in 
GeV) $v^{\prime\prime}_1=42$, $v^{\prime\prime}_2=1.4$, $v^{\prime\prime}_3=0.062$, 
$v^{\prime\prime}_{\phi_1}=2.552$, $v^{\prime\prime}_{\phi_2}=0.0074$, and $v^{\prime\prime}_{\phi_3}=
11.83$ we obtain the masses: $m_d=0.006,\,m_s=0.1,m_b=4.2$. For the $u$-type quarks we use, $a_u=a_d$, but now
$h_u=h^\prime_u=3$, and $v^\prime_1=53.8,\,v^\prime_2=0.4,\,v^\prime_3=0.0001$ and 
$v^\prime_{\phi_1}=0.1,\,v^\prime_{\phi_2}=124,\,v^\prime_{\phi_3}=0.082$, which give $m_u=0.0025$, 
$m_c=1.27$, and $m_t=171.2$. With these VEVs and parameters we obtain $V^{D,U}_L$ matrices given, 
according with the definition above, the mixing matrix
\begin{equation}
\vert V_{CKM}\vert =\left(\begin{array}{ccc}
0.97747\, & 0.21105& 0.00148\\
0.21083&\, 0.97673 &0.03923\\
0.00972\, & 0.03803 & 0.99923\\
\end{array}\right).
\label{ckm}
\end{equation}
Comparing the elements of the matrix above with experimental data in Section~11.2 of 
PDG~\cite{pdg}, we see that all the matrix elements, but $us$ and $ub$ entries, are within the 
level of confidence (1-2)$\sigma$. However, this is enough to show that a realistic mixing matrix 
may arise from the quark mass matrices given in Eqs.~(\ref{us}) and (\ref{ds}). A more rigorous and 
detailed analysis is being done and will be presented later.

The suppression of the FCNCs coupled to neutral Higgs bosons can be obtained not naturally 
in the sense of Ref.~\cite{gw}, but at least as a reasonable fine tuning among the mixing matrix
in the neutral scalar sector, $U_{ij}$, or by the mass of heavy Higgs scalars. For instance, consider 
the neutral kaons mass difference, $\Delta M_K \sim \zeta^i_{sd}f^2_Km_K(\sum_j\vert U_{ij}
\vert^2/m^2_j)$ where $\zeta^i_{sd}=h^2_d\,\vert\sum_a(V^D_L)^*_{as}(V^D_R)_{ad}\vert^2\,,$
with  symmetry eigenstates, say $H^{\prime\prime}_i$ ($i$ fixed), and physical scalars 
$h_j$ with mass $m_j$. In order to be consistent with data $\zeta^i_{sd}$ must be smaller 
than the contribution of the SM: i.e., smaller that $G^2_Fm^2_c (V^*_{cd}V_{cs})^2/16\pi^2=0.5\times
10^{-8}\textrm{GeV}^{-2}$ (for simplicity we have taken into account only 
the $c$ quark contribution and neglect the QCD and $m^2_c/M^2_W$ corrections). Notice that the sum over 
the contributions of each $i$ is made coherently. The expression above is valid for any scalar heavier 
than $m_K$. For instance, $h^2_d\vert\sum_a(V^D_L)^*_{as}(V^D_R)_{ad}\vert^2\sim10^{-4}-10^{-6}$ and 
$\vert U_{ij}\vert^2/m^2_j\sim(10^{-2}-10^{-4})\textrm{GeV}^{-2}$ we have the appropriate FCNC suppression. 
These constraints are not a large fine tunings since it can be satisfied with a large set of unitary
matrices and/or the neutral scalar mass spectrum and mixing.
Recall that there is no more freedom to redefine the quark phases since they all have already 
been used to put the CKM matrix in the usual form, hence $V^D_R$ has five physical phases. 
The case of FCNCs in the $u$-quark sector involve the matrices $V^U_L$ and $V^U_R$. 

On the other hand, it would be interesting if we can show that appropriate hierarchies among
the VEVs arises from an analysis of the scalar potential~\cite{zee0}. We have verified that, 
under reasonable assumptions on the parameters in the scalar potential, this is indeed the case. 
For the sake of simplicity we will consider the scalar potential involving only the scalar 
doublets that couple to quarks: the $A_4$ triplets $H^\prime,\Phi^\prime$ for $u$-like quarks, and
$H^{\prime\prime},\Phi^{\prime\prime}$ for $d$-like quarks. Denoting $X_u=H^{\prime\dagger}H^\prime$ 
and $Y_u~=~\Phi^{\prime\dagger}\Phi^\prime$, $X_d=H^{\prime\prime\dagger}H^{\prime\prime}$ and
$Y_d~=~\Phi^{\prime\prime\dagger}\Phi^{\prime\prime}$ we write the most general scalar
potential invariant under the SM and $A_4\otimes Z_3\otimes Z^{\prime}_3 \otimes Z^{\prime\prime}_3$
symmetries as:
\begin{eqnarray}
\mathcal{V}&=&\mu^2_p[X_p]_{1} \!+\! \mu^{\prime\,2}_p[Y_p]_{1} \!+\!
\mu^{2}_{\zeta}\vert\zeta\vert^2 \!+\! \lambda_{\zeta}\vert\zeta\vert^4 \!+\!
(\mu^2_{H^\prime\Phi^{\prime\prime}}[H^\prime\epsilon\Phi^{\prime\prime}]_{1}
\!+\! \mu^{2}_{H^{\prime \prime}\Phi^{\prime}}
[H^{\prime \prime} \varepsilon \Phi^{\prime}]_{1}
\nonumber \\ &+&
 f [\Phi^{\prime \dag} H^{\prime}]_{1}\zeta \!+\!
 f^\prime [\Phi^{\prime \prime \dag} H^{\prime \prime}]_{1}\zeta^{*}
\!+\! f_{H}^{\prime}[H^{\prime \prime} \varepsilon H^{\prime}]_{1} \zeta \!+\!
 f_{\Phi}^{\prime}[\Phi^{\prime \prime} \epsilon \Phi^{\prime}]_{1}\zeta^{*} \nonumber \\ &+&
\lambda^{pp^\prime}_{mn}[X_p]_m[X_{p^\prime}]_n+
K^{pp^\prime}_{kl}[Y_p]_k[Y_{p^\prime}]_l
\!+\! h^{pp^\prime}_{\alpha\beta}[X_p]_\alpha[Y_{p^\prime}]_\beta
+g_{st}[H^\prime\epsilon\Phi^{\prime\prime}]_s[H^\prime\epsilon
\Phi^{\prime\prime}]_t \nonumber \\ &+& g^{\prime}_{st}[H^{\prime \prime} \varepsilon
\Phi^{ \prime}]_{s}[H^{\prime \prime} \varepsilon \Phi^{ \prime}]_{t}+
g_{st}^{\prime \prime}([H^\prime\epsilon\Phi^{\prime\prime}]_{s}[H^{\prime \prime}
 \varepsilon \Phi^{\prime}]_{t} )
+ H. c.).
\label{potential}
\end{eqnarray}

We have omitted summation symbols and used the notation $p,p^\prime=u,d$;
$mn=11,1^\prime1^{\prime\prime},AA$; $kl=11,1^\prime1^{\prime\prime},BB$;
$\alpha\beta=11,1^\prime1^{\prime\prime},AB$; and
$st=11,1^\prime1^{\prime\prime},AA,BB,AB$. For example,
$\lambda^{pp^\prime}_{mn}[X_p]_m[X_{p^\prime}]_n =
\lambda^{pp^\prime}_{11}[X_p]_1[X_{p^\prime}]_1 +
\lambda^{pp^\prime}_{1^{\prime} 1^{\prime \prime}}[X_p]_{1^{\prime}}
[X_{p^\prime}]_{1^{\prime \prime}} +  \lambda^{pp^\prime}_{AA}[X_p]_A
[X_{p^\prime}]_A$ $+ \cdots$, and each term having contributions
$p, p^{\prime} = u,d$. Considering the Eq.~(\ref{potential}) we obtain the
constraint equations arising from the conditions $\partial \mathcal{V}/\partial
v^{\prime\prime}_{i} = 0, \dots$. For simplicity, in the following analysis it will
be also assumed that the VEVs are all real. In this case the condition
$1+\omega+\omega^2=0$ implies that in the constraint equations the coupling
constants will appear as the sum of some of the coupling constants in
Eq.~(\ref{potential}). The inclusion of all the scalar fields of the model 
and also the constraints coming from the condition  $\partial^2 \mathcal{V}/\partial 
v^{\prime\prime}_i\partial v^{\prime\prime}_j>0,\dots$ will be shown elsewhere. We have 
obtained the constraint equations under the following conditions: 
$v^{\prime\prime}_i=v^{\prime\prime}_{\phi_i},\;\forall i$;  and $\lambda^{dd}_{mn}
=K^{dd}_{kl}$, $\forall mn;kl$; and verified that, among the solutions to those
equations, it is possible to have the following hierarchy: $v^{\prime\prime}_1\gg 
v^{\prime\prime}_3 \gg v^{\prime\prime}_2$. However for generating the quark masses, as 
we have shown above, it is necessary that 
$v^{\prime(\prime\prime)}_1\gg v^{\prime(\prime\prime)}_{\phi_1}$. Notwithstanding, 
it is interesting that such hierarchies can arise from the model, and we hope that a more 
detailed treatment of the scalar potential will produce a more realistic hierarchies among 
the VEVs. 

As in any model with broken discrete symmetries there are potential troubles
with domain walls~\cite{walls}. However, this issue can be solved by allowing
soft terms in the scalar potential violating these symmetries.

In conclusion, the mass matrices obtained, Eqs.~(\ref{leptons}), (\ref{us}), (\ref{ds}),
and (\ref{neutrinos}), arising because of the symmetries of the model, give appropriate
insight concerning the solution of the flavor problem generating at the same time a realistic 
$V_{CKM}$ matrix. Of course, it is necessary to
explain how these symmetries are realized from a more fundamental theory.
Our scheme has some similarities with the ``private Higgs" of Porto and Zee~\cite{porto}
in which there is one Higgs per fermion and a general mass matrices. However, we have more than 
one Higgs per fermion and the mass matrices have a specific texture of zeros.

We thank an anonymous referee for useful comments. ACBM was fully supported and VP partially 
supported by CNPq.



\end{document}